\documentclass[a4paper]{article}

\usepackage{spconf}

\usepackage{amsmath,graphicx}
\usepackage{multicol,multirow,array}
\usepackage{tikz}
\usepackage[position=bottom]{subfig}
\usetikzlibrary{arrows,automata}

\title{Unified End-to-End Speech Recognition and Endpointing for Fast and Efficient Speech Systems}

\name{Shaan Bijwadia, Shuo-yiin Chang, Bo Li, Tara Sainath, Chao Zhang, Yanzhang He}
\address{Google, Inc.}

\copyrightnotice{978-1-6654-7189-3/22/\$31.00~\copyright2023 IEEE}
\begin{document}
\ninept
\maketitle

\begin{abstract}

Automatic speech recognition (ASR) systems typically rely on an external endpointer (EP) model to identify speech boundaries. In this work, we propose a method to jointly train the ASR and EP tasks in a single end-to-end (E2E) multitask model, improving EP quality by optionally leveraging information from the ASR audio encoder. We introduce a ``switch" connection, which trains the EP to consume either the audio frames directly or low-level latent representations from the ASR model. This results in a single E2E model that can be used during inference to perform frame filtering at low cost, and also make high quality end-of-query (EOQ) predictions based on ongoing ASR computation. We present results on a voice search test set showing that, compared to separate single-task models, this approach reduces median endpoint latency by 120 ms (30.8\% reduction), and 90th percentile latency by 170 ms (23.0\% reduction), without regressing word error rate. For continuous recognition, WER improves by 10.6\% (relative). \\

\end{abstract}

\section{Introduction}
\label{sec:intro}
\vspace*{-0.5ex}

End-to-end (E2E) approaches to automatic speech recognition (ASR) have been the subject of growing interest in recent years, and ASR is now a prominent input modality for many products, including digital personal assistants, smart speakers, and smartphone-based applications \cite{Schalkwyk2010, FERNANDES2021180}. E2E models outperform conventional ASR systems by integrating multiple tasks (acoustic modeling and language modeling) into a single model and training them jointly \cite{Graves12}. In this work, we continue this integration by incorporating another model which is typically trained separately, the endpointer model (EP).

The EP model is typically trained as small standalone model (e.g., \cite{chang2017endpoint, maas2018}), which assists recognition by generating two types of signals, voice activity detection (VAD) and end-of-query (EOQ) detection. First, the VAD task is to classify each frame according to whether it contains speech or silence. This signal is then used for ``frame filtering," i.e. discarding non-speech frames from the input before passing it to the much larger ASR model. For streaming ASR systems (which operate on audio frames in real-time as they are received), this reduces computation by allowing the system to ``skip" unnecessary frames. This is particularly important for continuous-query tasks, such as voice dictation, where recognition may run for an arbitrarily long period and computational savings accumulate over time.

The other primary task for the EP model is EOQ prediction, which is specific to ``short-query" speech tasks, such as digital assistants or interactive voice response (IVR) applications (e.g. ``Play Bruno Mars."). The task is to predict when the user is done speaking, at which point the system should close the microphone and generate a response \cite{chang2017endpoint}. High-quality EOQ detection is critical to reducing system latency, since a response typically will not be generated until the system recognizes that the user has finished speaking. For voice recognition systems, user-perceived latency (UPL) is an important factor for a satisfactory experience \cite{ward2005}. While complete speech recognition systems are complex and many individual components impact UPL, the quality of EP prediction is the largest single determinant \cite{Shangguan2021DissectingUL}.

Integrating the ASR and EP models is a natural next step in the progression towards E2E recognition. E2E training can confer two main benefits for this application. First, recognition and endpointing quality may improve, as joint training is likely to lead to better quality models by forcing the model to learn representations that generalize well across related tasks. Secondly, E2E training of ASR and EP reduces the infrastructure burden of building speech recognition systems, since only a single model would need to be trained, deployed, and maintained.

Other E2E approaches to jointly train EP and ASR mostly focus on subsuming the EP task into the ASR task, which we term ``decoder-based" EOQ detection, since they require inference by the ASR decoder to produce an EOQ signal. Decoder-based approaches can be effective for EOQ detection \cite{Shangguan2021DissectingUL}, but maintaining an acoustic-based EOQ detector helps cover the long-tail of utterances that the decoder-based EOQ does not detect properly \cite{Shuoyiin19}. Additionally, ASR decoding can be susceptible to high latency during streaming recognition if frame batching into larger chunks is being imposed for computational efficiency \cite{Ryan19}. Also, frame filtering based on VAD prediction is impossible using only a decoder-based signal. Therefore, the proposed method aims to augment the capabilities of a decoder-based EOQ detection with acoustic-based endpointing in a fully E2E setup. Best results for EOQ detection are obtained when combining acoustic- and decoder-based methods \cite{maas2018, Shangguan2021DissectingUL, Shuoyiin19}.

Therefore, our proposed method integrates the EP model into the audio encoder of an E2E ASR model to produce an acoustic-based EP prediction. A straightforward way to accomplish this is to share layers with the ASR encoder, a technique known as ``hard parameter sharing" \cite{ruder2017}. Li et al. \cite{Li2021LongRunningSR} previously explored training the VAD task on top of the lowest layers of an ASR model, with good results. However, directly applying this technique would cause issues when deployed for inference. EP models are typically kept as small as possible in order to be computationally efficient, since they must produce a VAD prediction for every audio frame (for frame filtering) \cite{braun2021}. Since ASR models are much larger than EP models, layer sharing may not be computationally viable. For example, each encoder layer in the ASR model described in \cite{Sainath2022} is approximately 32 times larger than the endpointer model described in \cite{chang19unified}. Running that single ASR layer for a VAD prediction on every audio frame would require a major increase in computation. Addressing this concern is the motivation for the novel architecture we propose here.

The key insight for our proposed method is that while frame filtering using a VAD signal is subject to a strict computational constraint, the EOQ task is free to leverage computation from the ASR model, since its signal is required \textit{only while speech is ongoing}. It is therefore desirable to have a small EP model that can operate on audio input directly, but also operate on the latent representation from the ASR system when it's available. During inference, the EP can be fed audio frames until it detects speech onset, at which point the audio will be fed to the ASR model, ``activating" the shared layers; the EP model can then be fed the latent representations from the shared layers until speech offset is detected. This scheme is most desirable in the short-query scenario, where improvements to EOQ prediction directly impact UPL, since the system can more quickly recognize that the user is finished speaking. We train this EP model by introducing the novel ``switch" connection: for each training example, the input to the EP is randomly chosen between the two possible inputs, the audio frames and the latent representation from the shared layers. Thus the model learns to expect either type of input, and can produce a prediction accordingly.

We evaluate the proposed method on real-world testsets, evaluating EOQ performance on short-query utterances and frame filtering performance on voice dictation data. We find that the proposed method reduces median EOQ detection latency improves by 120ms, a 30.8\% reduction, and 90th percentile latency by 170ms, a 23.0\% reduction, with no regression in word error rate (WER). Because EOQ detection directly affects user experience by ending recognition and generating a response faster, this is a substantial improvement to UPL. Further, we demonstrate on a continuous-query voice dictation set that frame filtering performance does not suffer relative to baseline; in fact, the word-error rate for the multitask ASR model improves, possibly by integrating the EP target signal into its acoustic understanding.
\section{Related Work}
\label{sec:related}
\vspace*{-0.5ex}
Early work on VAD relied on hand-crafted acoustic features, such as zero-crossing rate, energy ratio, and signal periodicity \cite{zaydin2019ExaminationOE, sakhnov2009}. Later, supervised machine-learning methods, including Hidden Markov Models (HMM) and Gaussian Mixture Models (GMM), were also shown to be effective \cite{Wu2018ASG}. Recent approaches have focused on deep neural network (DNN) structures, and long short-term memory (LSTM) modules in particular have been shown to perform well \cite{Eyben2013ReallifeVA, chang2017endpoint, Tong2016ACS}. As mentioned above, Li et al. implement VAD prediction by placing a fully-connected layer on top of the convolutional neural network (CNN) encoder of an ASR model \cite{Li2021LongRunningSR}.

For the reasons explored in Section \ref{sec:intro}, E2E integration of the EP and ASR has been of recent interest, and several decoder-based approaches have been explored. Yoshimura et al. \cite{Yoshimura2020EndtoEndAS} implement VAD by regarding the blank tokens in a CTC-based ASR model as the non-speech region. Chang et al. \cite{Shuoyiin19} perform EOQ detection by augmenting the ASR search space with an end-of-sentence token. Lu et al. extend the technique proposed in \cite{Shuoyiin19} to multi-speaker recognition \cite{Lu2022EndpointDF}. As stated in Section \ref{sec:intro}, decoder-based methods complement but cannot fully replace acoustic EP signals.

Our study builds on these prior works and offers several new contributions. While prior work in joint ASR and EP training in \cite{Li2021LongRunningSR} focuses solely on the VAD task, we perform both VAD and EOQ tasks; leveraging the ASR encoder for E2E prediction is a novel approach to acoustic-based EOQ. Additionally, to the authors' knowledge, the ``switch" connection is an entirely novel solution to the problem of differing computational constraints for frame filtering versus EOQ detection. This is a critical technique for real-world deployment of an E2E ASR and EP model, as it must meet the constraints of a real-time speech recognition system.

\section{Model}
\label{sec:model}
\vspace*{-.5ex}

\begin{figure}[t]
  \centering
  \includegraphics[width=\columnwidth]{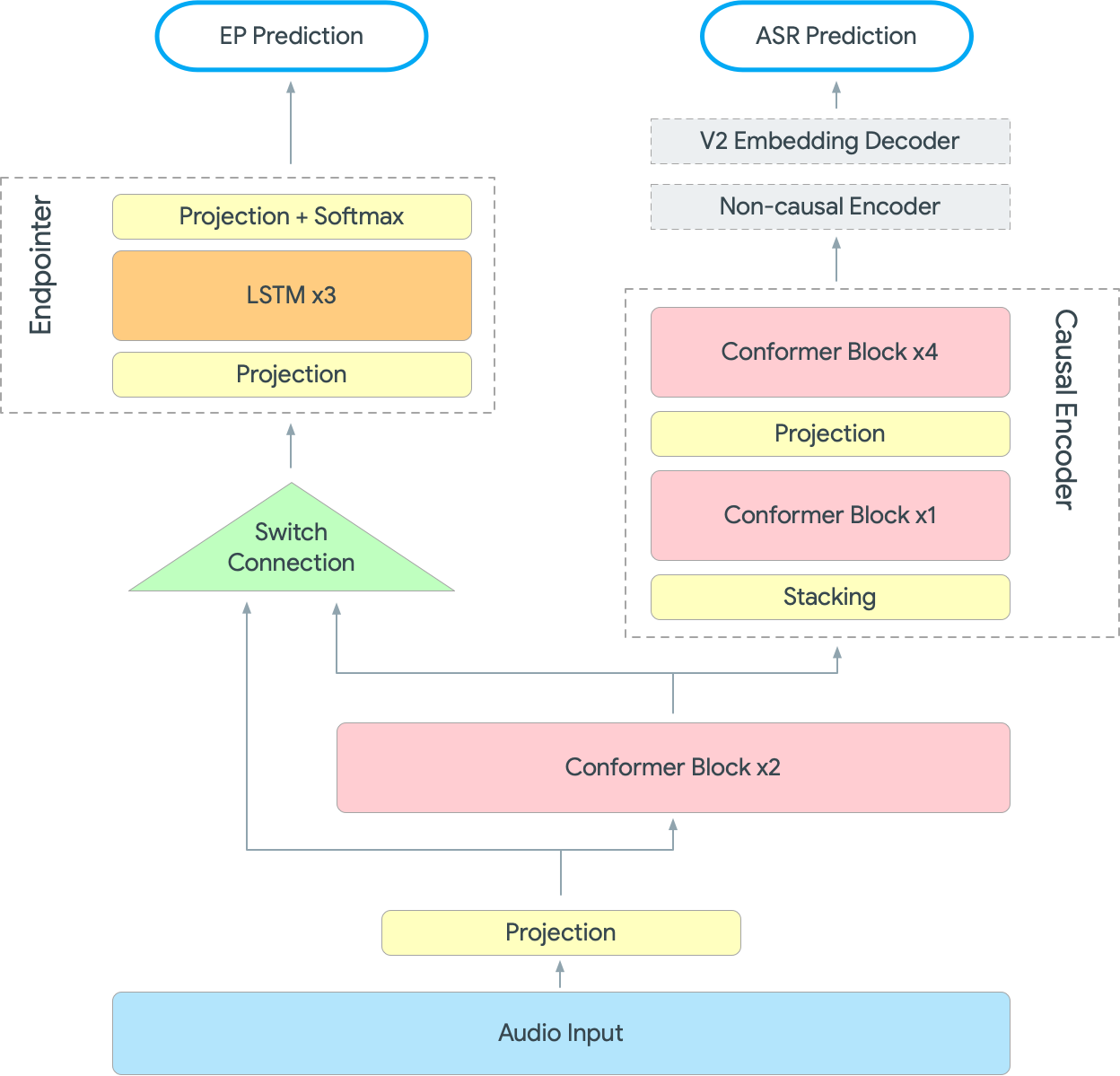}
    \vspace{-0.1in}
  \caption{\textbf{Model Architecture} \: Multitask ASR and EP model, featuring shared encoder layers and a switch connection. During training, the switch connection randomly chooses between two inputs with equal probability for each training example.}
  \vspace{-0.1in}
  \label{fig:architecture}
\end{figure}

\subsection{Endpointing Task}
\label{subsec:ep-task}

Similar to \cite{chang19unified}, the endpointing task is modeled as a frame-level classification task, where each frame is classified as one of four speech classes:
\begin{gather}
    \mathcal{S} = \biggl\{\hspace{1em}
    \begin{aligned}
        &\text{speech},\\
        &\text{initial silence},\\
        &\text{intermediate silence},\\
        &\text{final silence}
    \end{aligned}
    \hspace{1em}\biggl\}
\end{gather}
Within an utterance, all frames before speech begins are labeled ``initial" silence, and all frames after the final speech segment are labeled ``final;" all other segments are labeled ``intermediate." Because this method is intended for streaming speech recognition scenarios, we allow the EP model to condition its predictions on previous frames. Formally, for each audio frame at time $t$ the model predicts: 
\begin{gather}
    P(S_t | X_{0:t})
\end{gather}
where $X$ is the sequence of input frames, and $S$ is the sequence of frame-wise speech class labels, $S_t\in\mathcal{S}$.

We therefore achieve a categorical distribution over $\mathcal{S}$ for every frame; at inference time, we derive the VAD signal for each frame by applying a threshold to the predicted probability $P(S_t=\text{speech} | X_{0:t})$, and likewise for the EOQ signal using $P(S_t=\text{final silence} | X_{0:t})$.

We optimize the EP task using standard cross-entropy loss
\begin{gather}
    \mathcal{L}_{\text{EP}} = \text{CE}(\hat{S}, S)
\end{gather}
where $\hat{S}$ is the predicted class label probabilities for a given utterance.

\subsection{ASR Task}

E2E ASR models derived from the RNN-T structure \cite{Graves12} involve an audio encoder and prediction network, roughly analogous to an acoustic model (AM) and language model (LM), respectively, in classical ASR systems. Information from both networks are joined together to produce a prediction $P(Y|X)$, where $Y$ is a tokenized representation of the transcript. This is optimized using RNN-T loss, described more fully in \cite{Graves12}. Briefly: the ASR model is trained to output sequences containing valid wordpiece tokens or special ``blank" tokens that are ignored; predictions containing blanks are called ``alignments." The training loss is:
\begin{gather}
    \mathcal{L}_{\text{ASR}} = \log P(Y|X) = \sum_{\text{A}: \mathcal{B}(A) = Y} P(A|X)
\end{gather}
where $\mathcal{B}$ is a function mapping an alignment $A$ to valid output sequences by removing all instances of the blank symbol \cite{Graves06}.

\subsection{Multitask Model and Switch Connection} \label{arch}
In the proposed model, ASR and EP are trained jointly. We first modify the training loss to be a weighted average of the ASR and EP loss functions:

\begin{gather}
    \mathcal{L}_{\text{multi}} = \lambda \mathcal{L}_{\text{ASR}} + (1 - \lambda) \mathcal{L}_{\text{EP}}
\end{gather}
where $\lambda\in [0, 1]$ is a hyperparameter defining the training weight given to the ASR task.

The first two conformer layers of the ASR's audio encoder are shared with the EP via hard parameter sharing \cite{ruder2017}. The EP model may consume either the audio frames directly or the latent representation from the shared layers. During training, the input given to the endpointer is determined stochastically per utterance by selecting the audio frames or the latent features with equal probability. We term this a ``switch" connection. A diagram of this model is shown in Figure \ref{fig:architecture}.

\begin{figure}[t]
\centering
\vspace{-1em}
\subfloat[State machine for short queries, e.g. ``Play Bruno Mars."]{
    \begin{tikzpicture}[>=stealth',shorten >=1pt,auto,node distance=3cm]
        \ninept
        \node[initial,state] (audio)      {EP Only};
        \node[state]         (latent) [right of=audio]  { ASR + EP};
        \node[state,accepting] (end) [right of=latent] {End};
      
        \path[->] (audio)
            edge [loop above] node {} (audio)
            edge [bend right=30, swap] node {$P(\text{speech}) > \theta_{\text{VAD}}$} (latent)
        (latent)
            edge [loop above] node {} (edge)
            edge [bend right=14, swap] node {$P(\text{final}) > \theta_{\text{EOQ}}$} (end);
    \end{tikzpicture}
}
\vspace{-1em}
\subfloat[State machine for continuous recognition, e.g. ``To be, or not to be, that is the question..."]{
    \begin{tikzpicture}[>=stealth',shorten >=1pt,auto,node distance=4cm]
        \ninept
      \node[initial,state] (audio)      {EP Only};
      \node[state]         (latent) [right of=audio]  {ASR + EP};
      
      \path[->] (audio)
                    edge [loop above] node {} (audio)
                    edge [bend left] node {$P(\text{speech}) > \theta_{\text{VAD}}$} (latent)
                (latent)
                    edge [loop above] node {} (edge)
                    edge [bend left] node {$P(\text{speech}) < \theta_{\text{VAD}}$} (audio);
    \end{tikzpicture}
}
\caption{\textbf{Inference Logic} \: Finite state machine diagrams representing the logic determining which input is given to the EP during streaming recognition. While the EP predicts that the user is not speaking, inference is run on the EP only, using the audio frames as input (``EP Only"). When speech is detected by the EP, inference is run on both ASR and EP, using the ASR latent features as input for the EP (``ASR + EP").}
 \label{fig:FSM}
\end{figure}
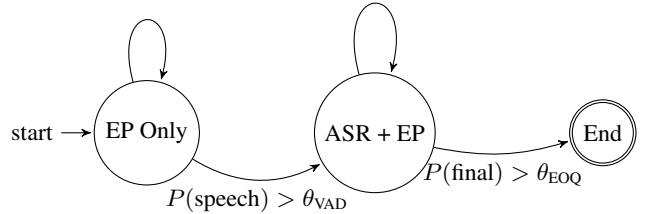
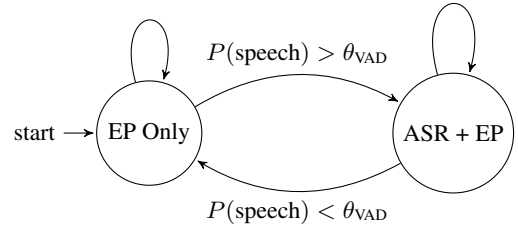

During inference, the switch connection is replaced with an input logic determined by the EP prediction for the previous frame. We maintain two thresholds, $\theta_{\text{VAD}}, \theta_{\text{EOQ}} \in [0, 1]$, which are hyperparameters. We say the EP predicts speech if $P(\text{speech}) > \theta_{\text{VAD}}$, and predicts EOQ if $P(\text{final silence}) > \theta_{EOQ}$. We begin by feeding audio frames to the EP, without passing them down to the ASR model. Then, when speech onset is detected, audio frames are passed to the ASR model, and the EP model performs inference with the ASR latent features as input.
The subsequent logic differs depending on the type of recognition. In continuous recognition, we simply return to the non-speech state once the EP-predicted speech posterior drops below $\theta_{\text{VAD}}$. In the short query case, we wait for the system to declare EOQ, using a combination of signals from the acoustic- and decoder-based EOQ detectors, at which point we end recognition (see \S\ref{subsec:eoq}). This logic is illustrated as a finite state machine diagram in Figure \ref{fig:FSM}.

\section{Experimental Setup}
\label{sec:expts}
\vspace*{-.5ex}

\subsection{EP Model}
Input to the endpointer model is passed through a fully-connected projection layer, which feeds into a 3-layer stack of 128-dim LSTM blocks \cite{HochreiterSchmidhuber97}, similar to \cite{maas2018, chang19unified}. The output from the LSTM encoder is fed to another fully-connected layer, which projects the 128-dim encoding to a 4-dim output vector. These values are passed as logits through a softmax function to obtain probability predictions for each frame.

\subsection{ASR Model}
As our ASR model, we use a $\sim$150M parameter streaming cascaded conformer-transducer (Conf-T) \cite{gulati2020conformer, Sainath2022}, which features a causal conformer encoder with 7 layers and attention dimension 512, and a $V^2$ embedding decoder (i.e., the prediction network computes embeddings based on the two most recent non-blank tokens). As this is a streaming model \cite{Zhang20}, the causal encoder is given only left-context during recognition. The encoder contains a time-reduction stacking layer after the second conformer layer, which down-samples the input by a factor of 2.

As this is a cascaded-encoder setup \cite{arun21cascade}, outputs from the causal encoder are passed to a second encoder which receives limited right context, composed of 6 conformer layers with dimension 384. The $V^2$ embedding decoder two produces predictions based on both the causal and non-causal encoder features; the word-error rates (WER) reported in Section \ref{s:results} are obtained from the non-causal pathway. This model also features an E2E-EP \cite{Shuoyiin19}, wherein the ASR search space is augmented with an $\texttt{</s>}$ token that signals EOQ (see \S\ref{subsec:eoq}).

\subsection{Training Setup}
The acoustic features used for all experiments are 128-dim log-mel features, computed in 32ms windows every 10ms. Each frame is stacked with three previous frames to produce a 512-dim vector and downsampled to 30ms, and concatenated with a 16-dim one-hot domain ID vector.

Our training dataset contains $\sim$400k hours of English utterances sampled from multiple domains, including voice-search tasks, far-field environments, telephony conversations, and audio drawn from longform videos. All utterances are anonymized and hand-transcribed, and augmented using room simulations and artificial noise with signal-to-noise ratios (SNR) between 0db and 30db. SpecAugment \cite{Park2019} is applied for regularization. Training labels for the endpointer are generated by running a forced word alignment based on the transcription label \cite{moreno1998recursive}. Once speech and non-speech regions are identified, the first non-speech segment is classified as ``initial silence," the last one as ``final silence," and all others as ``intermediate silence." All experiments are implemented in TensorFlow \cite{AbadiAgarwalBarhamEtAl15} and the Lingvo toolkit \cite{shen2019lingvo} on 8x8 slices of Tensor Processing Units \cite{JouppiYoungPatilEtAl17}.

The baseline used for comparison uses single-task versions of the above models, trained separately and combined into a complete speech recognition system. The single-task ASR model is trained for 600k steps, with a global minibatch size of 4096 utterances using the Adam optimizer \cite{KingmaBa15}. All ASR training includes FastEmit \cite{yu21fastemit} to encourage the model to emit hypotheses with low latency. The single-task EP model is trained for 300k steps, as the model converges much more quickly. For the multitask ASR and EP models, we set $\lambda=0.98$, and train with the same parameters described above for single-task ASR models.

We are aware of the sensitive nature of the  speech recognition research and other AI technologies used in this work. All training data used in this work abides by Google AI Principles \cite{googleaiprinciples}.

\subsection{Declaring EOQ}
\label{subsec:eoq}
For short-query utterances, recognition may be ended by either acoustic- and decoder-based EOQ detection. The proposed method is considered the acoustic-based EOQ detector. As explained in \S\ref{subsec:ep-task}, acoustic EOQ is predicted if $P(\text{final silence}) > \theta_{\text{EOQ}}$. Decoder-based EOQ detection is performed by augmenting the ASR target vocabulary with a special token $\texttt{</s>}$ indicating the predicted end of speech \cite{Shuoyiin19, li2020towards, Zhao2021PreventingEE}. The ASR model predicts EOQ if the top beam contains that token with prediction cost lower than a hyperparameter $\theta_{\texttt{</s>}}$. EOQ can be declared by either the acoustic- or decoder-based signal, whichever comes first. Once EOQ is detected, we impose a short mandatory waiting period $w$ to further avoid deletion errors due to early cutoff. For each experiment, $\theta_{\text{EOQ}}$, $\theta_{\texttt{</s>}}$, and $w$ are swept using grid search on the evaluation set to optimize median and 90th percentile latency for a given WER (chosen here to be 5.8). \textit{All} systems evaluated in this work have both acoustic- and decoder-based EOQ detection enabled, since the end-to-end nature of the proposed system may affect both methods of endpointing.

\subsection{Evaluation}
We evaluate EOQ detection performance on a short-query dataset containing $\sim$14k anonymized and hand-transcribed far-field utterances of voice search traffic. The main metrics of interest are WER, and endpoint latency, defined as the amount of time between when the user stops speaking (ground truth EOQ) and the detected EOQ. The ground truth EOQ is estimated using a forced word alignment in the same fashion as the endpointer training labels are generated. We report the median and 90th percentile latency (EP50 and EP90, respectively).

We evaluate frame filtering performance on a continuous-query dataset containing $\sim$14k anonymized and hand-transcribed utterances of voice dictation traffic, which are broken up into speech segments of $\sim$10 seconds on average. We report WER, and the percentage of the utterance that is left after frame filtering, which measures on how many frames ASR inference is run (lower is better).

\section{Results}
\label{s:results}
\vspace*{-0.5ex}

\begin{table}[t]
    \centering
    \caption{\textbf{Short-Query Results} \:ASR and EP performance for the proposed multitask model in a complete speech recognition system. We report word-error rate (WER (\%)), and 50th/90th percentile endpoint latency (EP50, EP90 (ms)).}
    \vspace*{-1ex}
    \label{tab:short-results}
    \begin{tabular}{|c|l|c|c|c|}
            \hline
        Exp. & Method & WER & EP50 & EP90 \\ \hline \hline
        B1 & Separate Models &  5.8 & 390 & 740 \\ \hline
        E1 & Multitask Model & 5.8 & 320 & 610 \\
        E2 & \hspace{.5em} + 2 Shared Layers & 5.8 & \textbf{270} & \textbf{560}  \\
        \textbf{E3} & \hspace{.5em} \textbf{+ Switch Connection} & \textbf{5.8} & \textbf{270} & 570 \\
    \hline
    \end{tabular}
    \vspace*{-1ex}
\end{table}

\begin{table}[t]
    \centering
    \caption{\textbf{Continuous Results} \: Frame filtering performance for the proposed system, evaluated on a continuous query test set. We report word-error rate (WER (\%)), broken down by deletions, insertions, and subsitutions. We also report the percentage of audio remaining after frame filtering (Speech \%).}
    \vspace*{-1ex}
    \label{tab:continuous-results}
    \begin{tabular}{|l|c|c|c|c|c|}
            \hline
        Exp. & WER & del & ins & sub & Speech \% \\ \hline \hline
        B1 & 10.4 & \textbf{1.1} & 4.7 & 4.5 & \textbf{0.73} \\
        B1, no filter & 14.4 & 1.2 & 8.2 & 5.0 & 1.0 \\ \hline
        \textbf{E3} & \textbf{9.3} & 1.2 & \textbf{3.6} & \textbf{4.4} & \textbf{0.73} \\
        E3, no filter & 13.6 & 1.2 & 7.6 & 4.8 & 1.0 \\ \hline
    \end{tabular}
    \vspace*{-2ex}
\end{table}

\subsection{EOQ Detection}
We present results for a step-wise progression towards the proposed system in Table \ref{tab:short-results}. Our baseline (B1) configuration uses separately trained ASR and EP models, combined in a single speech recognition pipeline. The first experiment (E1) trains these two models jointly using the multitask loss function described in \S\ref{arch}, but \textit{without} any parameter sharing. This helps verify that the multitask learning task is effective, independent of other architectural changes. We then introduce parameter sharing (E2), placing the EP model downstream of the second conformer layer in the ASR audio encoder. At this stage, all input to the EP is passed through the first two conformer ASR layers for both VAD and EP -- this allows us to assess the quality impact of sharing parameters with the ASR model. This improves median and 90th percentile latency by 30.8\% and 23.0\%, respectively. Note that this is not a deployable model, since the VAD prediction would be too expensive for frame filtering.

Finally, we introduce the switch connection (E3), which is our proposed candidate model. The ideal case for this experiment is to match the quality of E2 -- we observe that training with a switch connection does not degrade the median latency, and only marginally increases 90th percentile latency.
\vspace*{-0.5ex}

\subsection{Frame Filtering}

We present the frame filtering performance of the our proposed E2E model on continuous-query utterances in Table \ref{tab:continuous-results}. The results compare the baseline (B1) and proposed system (E3), which are exactly the same models as listed in Table \ref{tab:short-results}. We evaluate both systems with frame filtering enabled and disabled, to demonstrate the effect of filering on WER. The results show that both systems are similarly effective at filtering, leaving 73\% of the utterance remaining after discarding silence frames. Note that a major source of word errors in both systems come in the form of insertions -- we observe that the E2E ASR models examined in this study are prone to hallucinating words during noisy non-speech parts of the audio. Therefore, frame filtering has the additional benefit of reducing insertion WER, by recognizing and filtering out non-speech frames. This is why applying the frame filter actually reduces WER.

Perhaps surprisingly, the ASR system in E3 has superior WER, and most of the relative reduction in WER is due to lower insertion errors. We speculate that in the multitask architecture, the shared encoder layers learn from the EP target about which frames contain speech. This makes the ASR results more resistant to the aforementioned hallucinations during periods of non-speech noise, even when the EP is inactive.

\begin{figure}[t]
  \centering
  \includegraphics[width=\columnwidth]{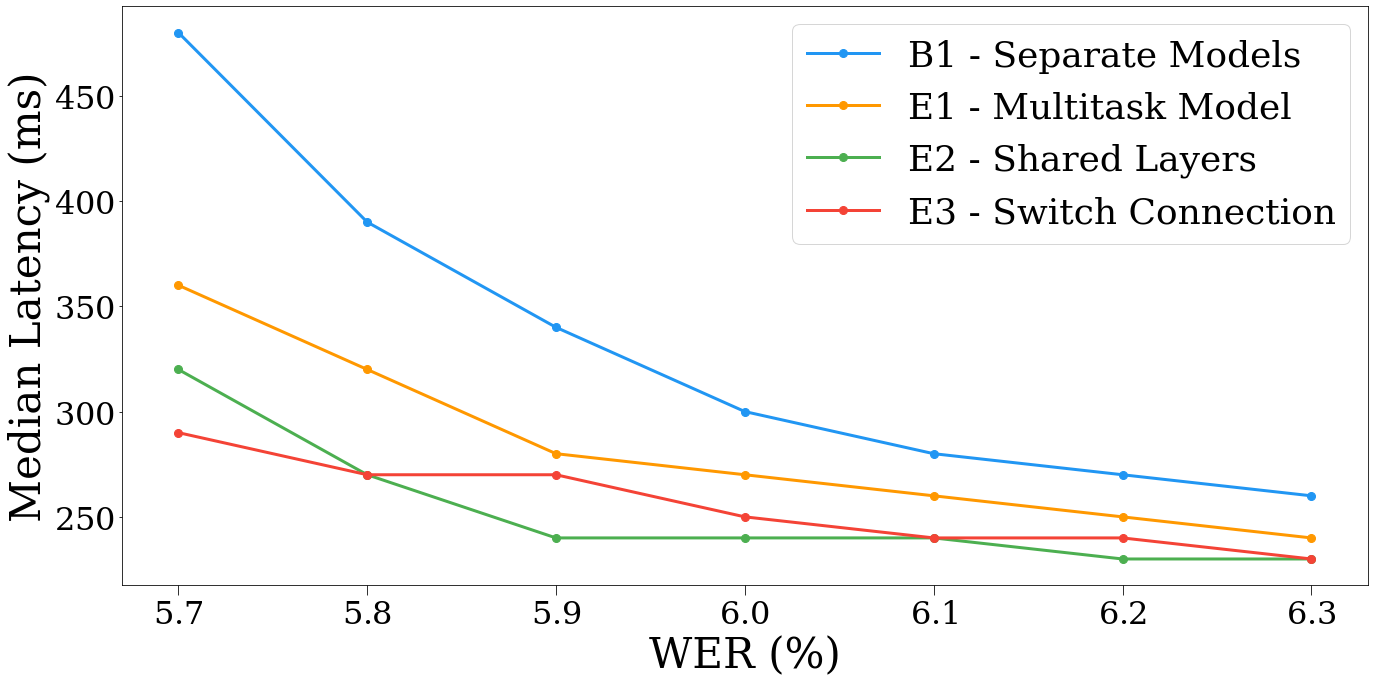}
  \caption{\textbf{WER vs. Median Latency} \: Response curves showing the optimal median latency for a given WER on the short query test set.}
  \label{fig:curves}
\end{figure}

\begin{figure}[t]
  \centering
  \subfloat[``What time is it?"]{
    \includegraphics[width=\columnwidth]{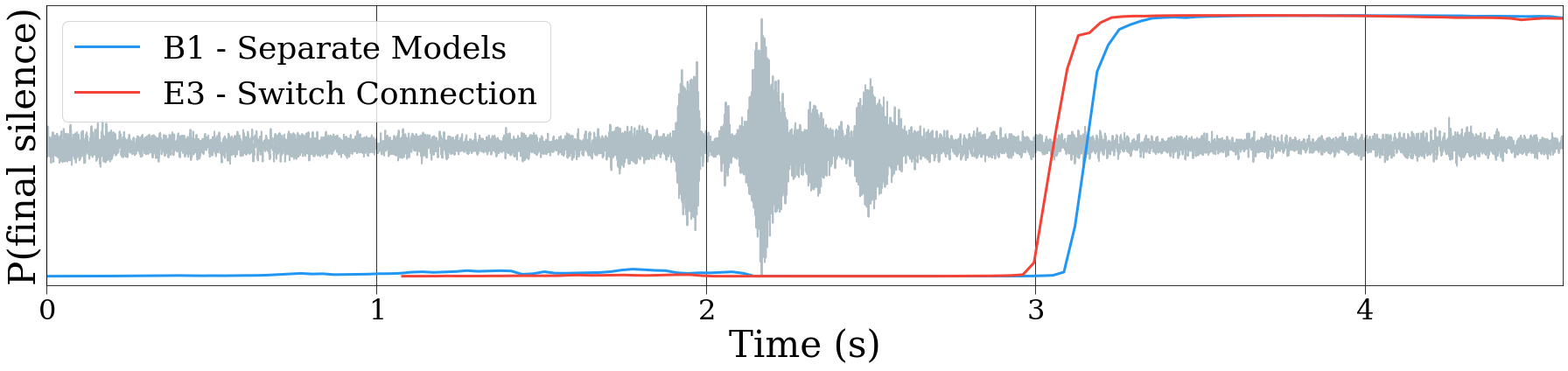}
  }\\
  \subfloat[``How many people speak German?"]{
    \includegraphics[width=\columnwidth]{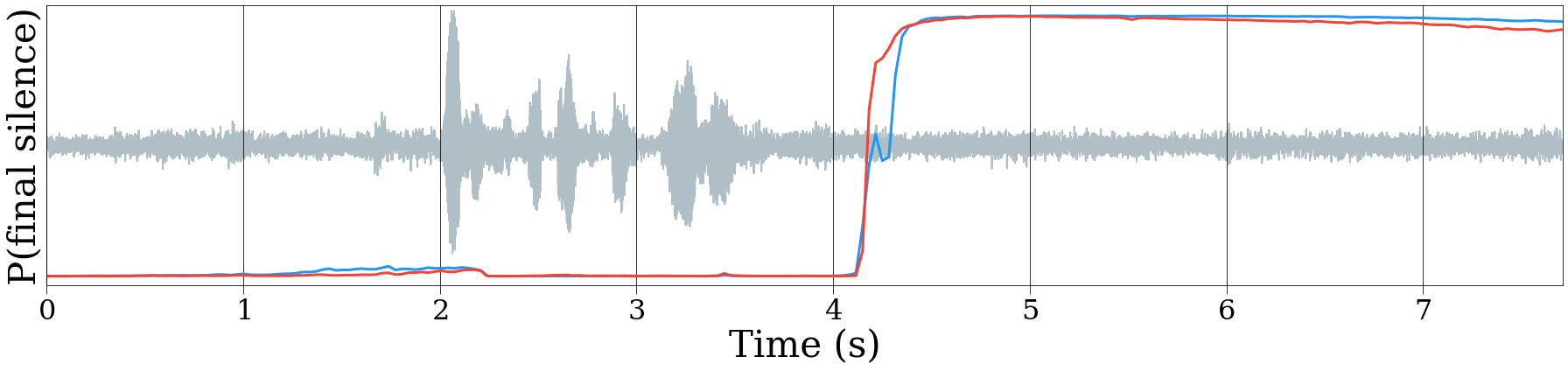}
  }\\
  \subfloat[``Open Maps."]{
    \includegraphics[width=\columnwidth]{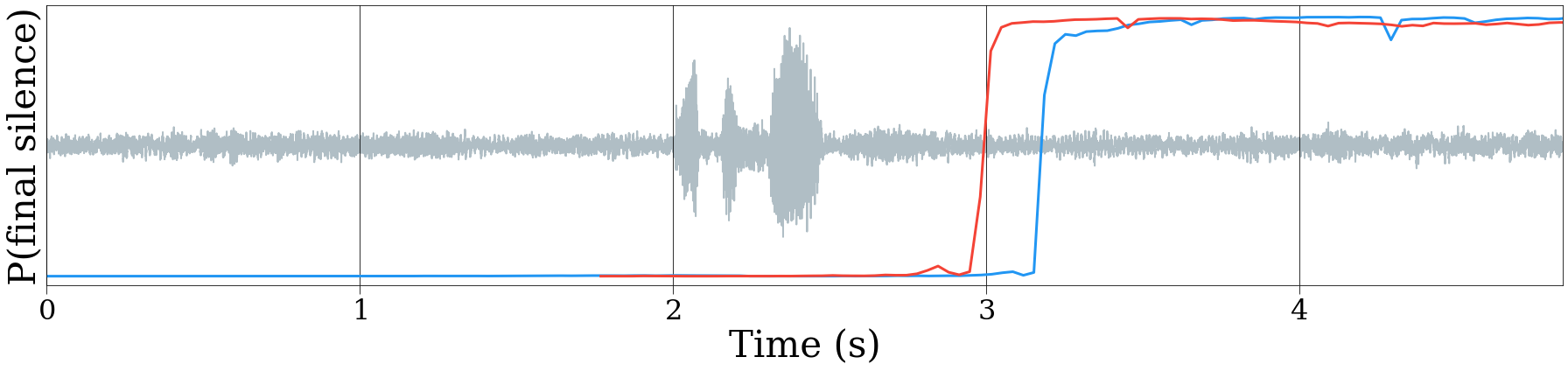}
  }
  \caption{\textbf{EOQ Posterior} \: Plots showing the predicted posterior $P(\text{final silence})$ during recognition on representative examples of human-spoken short queries.}
  \label{fig:posterior}
\end{figure}

\subsection{Analysis}

We now further analyze the performance of the proposed system. The EOQ task involves a tradeoff between WER and latency, since more aggressive EOQ detection may lead to deletion errors if EOQ is declared too early. Therefore, multiple operating points along this tradeoff curve are possible. To better visualize the performance of these systems across WERs, we plot the median latency for various WERs for the test set evaluated in Table \ref{tab:short-results}. We depict this plot in Figure \ref{fig:curves}; note that lower curves (toward the bottom right) are better.  All multitask experiments outperform B1 at every WER, demonstrating the effectiveness of joint learning. E2 offers a substantial advantage over E1, since a shared layers allow the EP to make higher quality EOQ decisions. We observe that introducing the switch connection in E3 offers similar performance across the range of WERs as E2. We also present visualizations of the EOQ posterior during recognition for short-query utterances, demonstrating the quicker recognition of the proposed system compared to baseline.

To better understand the speedup in EOQ latency, we plot the posterior of the predicted $P(\text{final silence})$ posteriors for the baseline (B1) and proposed (E3) systems over time during utterance recognition. Figure 4 provides three representative examples, which show that the proposed system often transitions between predicted states at a similar speed, but earlier in the utterance. Additionally, a common pattern, as pictured in example (b), is for the baseline system to momentarily lose confidence in declaring EOQ, causing a slight endpointing delay; the proposed system was not found to exhibit this behavior.
\section{Conclusions}
\label{sec:conclusions}
\vspace*{-.5ex}

 E2E modeling has rapidly become the preferred approach for building the high quality ASR models that underlie popular speech-based technologies. This work presents a multitask approach to jointly training the acoustic EP and ASR in a single E2E model, by sharing parameters between the ASR audio encoder and EP model. To enable low-cost frame filtering, we introduce the novel ``switch" connection, which trains the EP to accept either audio frames directly or the latent representation from the shared encoder layers. During inference, the EP can be used as a standalone model for VAD-based frame filtering while the user is not speaking, or as a high-quality EOQ predictor which leverages ongoing ASR computation. Applying this E2E architecture leads to substantial quality improvements in EOQ detection latency. WER for continuous recognition also improves, likely due to a better robustness against hallucinations due to non-speech noise. Additionally, unifying the ASR and EP tasks into a single model removes the infrastructure burden of maintaining two separate models. We recommend further research in the direction of shared acoustic understanding between ASR and EP tasks.

\bibliographystyle{IEEEbib}
\bibliography{refs}

\end{document}